\documentclass[pra,aps,twocolumn,floatfix,bibnotes,showpacs]{revtex4}

\usepackage{psfig}

\begin{document}

\title{Optimal discrimination of mixed quantum states involving
  inconclusive results}

\author{Jarom{\'{\i}}r Fiur{\'a}{\v{s}}ek}

\author{Miroslav Je{\v{z}}ek}

\affiliation{Department of Optics, Palack{\'y} University,
  17. listopadu 50, 77200 Olomouc, Czech Republic}

\date{\today}


\begin{abstract}

We propose a generalized discrimination scheme for mixed quantum states.
In the present scenario we allow for certain fixed fraction of inconclusive
results and we maximize the success rate of the quantum-state discrimination.
This protocol interpolates
between the Ivanovic-Dieks-Peres scheme and the Helstrom one.
We formulate the extremal equations for the optimal positive operator
valued measure describing the discrimination device
and establish a criterion for its optimality. We also devise
a numerical method for efficient solving of these extremal equations.
\end{abstract}

\maketitle

\section{Introduction}

The non-orthogonality of quantum states is one of the fundamental
features of quantum mechanics. It means that two different
quantum states cannot be perfectly discriminated in general.
This fact imposes limits on the processing
of quantum information. For instance, an unknown quantum state
cannot be copied \cite{Wootters_Zurek_1982,Dieks_1982}.
On the other hand, the non-orthogonality of the quantum states can
be advantageous. In particular, most of the quantum cryptographic
schemes rely on it \cite{Bennett_Brassard_1984,%
Bennett_Brassard_Mermin_1992,Bennett_1992}.
The principle of operation of these schemes is to employ several
non-orthogonal states to encode the bits of the secret key that
is sent from transmitter (Alice) to receiver (Bob). Since the
eavesdropper (Eve) cannot discriminate all these states perfectly,
any attempt to gain information on the key will increase the noise
of the transmission from Alice to Bob and thus Eve will be
unavoidably revealed.

Since one cannot discriminate non-orthogonal states perfectly,
a natural question arises what is the optimal approximate discrimination
scheme. Two different approaches to the problem have been proposed.
The first scenario, the ambiguous quantum state discrimination, has
been analyzed by Holevo, Helstrom and others
\cite{Holevo_1973,Yuen_Kennedy_Lax_1975,Helstrom_QDET,Hausladen_1994,%
Sasaki_Kato_Izutsu_Hirota_1998,Barnett_2001,Eldar_2001,%
Andersson_et_al_2002}.
They considered discrimination of $N$ mixed quantum states
$\rho_j$ that are generated by Alice with the a-priori
probabilities $p_j$,
\begin{equation}
\sum_{j=1}^N p_j=1, \qquad p_j>0.
\end{equation}
Bob performs a generalized quantum measurement described by the
$N$-component positive operator valued measure (POVM)
$\{\Pi_j\}_{j=1}^N$. If the outcome $\Pi_j$ is detected, then Bob
concludes that Alice sent the state $\rho_j$ to him. The optimal
POVM is defined as the POVM that maximizes the average success rate
of the Bob's guesses,
\begin{equation}
{P_{\rm S}} = \sum_{j=1}^N p_j {\rm Tr}[\Pi_j \rho_j].
\label{SR}
\end{equation}
An interesting alternative approach has been suggested by Ivanovic, Dieks,
and Peres (IDP) for the discrimination of two  pure
states \cite{Ivanovic_1987,Dieks_1988,Peres_1988,Jaeger95} and extended to $N$
linearly independent states by Chefles and Barnett
\cite{Chefles_1998,Chefles_Barnett_1998,Zhang_et_al_2001,Sun_2001}.
These states can be discriminated unambiguously, provided that
we allow for some fraction of the inconclusive results $P_{\rm I}$.
Recently, several other problems closely related to quantum-state
discrimination have been considered such as the discrimination of sets
of quantum states \cite{Herzog_2002,Zhang_2002} and the
quantum-state comparison \cite{Jex_2002,Sasaki_2002}.

Besides being a theoretically interesting problem, the quantum-state
discrimination has also found practical applications in quantum information
processing. In particular, the optimal unambiguous quantum-state
discrimination represents a simple and experimentally feasible
attack on the quantum key distribution protocols
\cite{Ekert_et_al_1994,Dusek_et_al_2000}.
Recently, some optimal discrimination POVMs have been realized
experimentally and the Helstrom as well as
the IDP bounds have been attained
\cite{Huttner_et_al_1996,Barnett_Riis_1997,%
Clarke_Chefles_Barnett_Riis_2001,Clarke_et_al_2001}.

The two above discussed scenarios can be considered as limiting
cases of a more general scheme that involves certain  fraction
of inconclusive results $P_{\rm I}$ for which we maximize
the success rate. The main feature of this scheme is that if we
allow for inconclusive results then we can improve the relative
(or re-normalized) success rate
\begin{equation}
{P_{\rm RS}}=\frac{P_{\rm S}}{1-P_{\rm I}}.
\label{RSR}
\end{equation}
In other words, with probability $P_{\rm I}$ Bob fails completely
and he cannot say at all which state was sent to him. However,
in the rest of the cases when he succeeds he can correctly  guess
the state with higher probability than if he would not allow for the
inconclusive results. For  pure linearly independent states this
generalized scenario was discussed in two recent papers
\cite{Chefles_Barnett_1997_JMO,Zhang_Li_Guo_1999}.

Here we extend the analysis to  general mixed states. At a first
sight, such an extension  may seem to be a bit problematic, because
it is known that one cannot unambiguously discriminate mixed states
(the reason is that the IDP scheme does not work for linearly
dependent states). Nevertheless, we shall show that the extension is
perfectly meaningful also for mixed states. Although we cannot reach
the limit of perfect unambiguous discrimination, we can improve the
${P_{\rm RS}}$ and we shall derive an upper bound on ${P_{\rm RS}}$ that
can be achieved for a given set of mixed states.

\section{Extremal equations for optimal POVM}

Let us begin with the formal definition of the problem.
We assume that the quantum state sent to Bob is drawn from
the set of $N$ mixed states $\{\rho_j\}_{j=1}^N$ with the a-priori
probabilities $p_j$. Bob's measurement on the state may yield
$N+1$ different results and is formally  described by the POVM
whose $N+1$ components satisfy
\begin{equation}
\Pi_j\geq 0, \quad j=0,\ldots, N, \qquad \sum_{j=0}^N \Pi_j= \openone,
\label{POVM}
\end{equation}
where $\openone$ is the identity operator.
The outcome $\Pi_0$ indicates failure and the probability of
inconclusive results is thus given by
\begin{equation}
P_{\rm I}= \sum_{j=1}^N p_j {\rm Tr}[\rho_j \Pi_0].
\label{PINCDEF}
\end{equation}
For a certain fixed value of $P_{\rm I}$ we want to maximize the
relative success rate (\ref{RSR}) which is equivalent to the
maximization of the success rate (\ref{SR}).
To account for the linear constraints
(\ref{POVM}) and (\ref{PINCDEF}) we introduce Lagrange multipliers
$\lambda$ and $a$ where $\lambda$ is Hermitian operator and $a$ is
a real number. Taking everything together we should maximize the
constrained success rate functional
\begin{equation}
\bar{P}_{\rm S}  = \sum_{j=1}^{N} p_j {\rm Tr}[\rho_j \Pi_j] -
\sum_{j=0}^N {\rm Tr}[\lambda \Pi_j] + a  \sum_{j=1}^N p_j
{\rm Tr}[\rho_j \Pi_0].
\label{RSRTILDE}
\end{equation}
We now derive the extremal equations that must be satisfied by the
optimal POVM. We expand the POVM elements in terms of their eigenstates
and eigenvalues, $\Pi_j=\sum_{k} r_{jk}
|\varphi_{jk}\rangle\langle\varphi_{jk}|$ and
vary (\ref{RSRTILDE}) with respect to $\langle \varphi_{jk}|$. After
some algebraic manipulations, we arrive at the extremal equations,
\begin{eqnarray}
(\lambda-p_j \rho_j) \Pi_j & = & 0, \qquad j=1,\ldots,N,
\label{EXTONE} \\
(\lambda-a \sigma)\Pi_0 & = & 0,
\label{EXTTWO}
\end{eqnarray}
where the operator $\sigma$ introduced for the sake of notational
simplicity reads
\begin{equation}
\sigma=\sum_{j=1}^N p_j\rho_j.
\label{SIGMA}
\end{equation}
From the constraint ${\rm Tr}[\sigma\Pi_0]=P_{\rm I}$  we can express
$a$ in terms  of $\lambda$,
\begin{equation}
a =P_{\rm I}^{-1} {\rm Tr}[\lambda \Pi_0].
\label{A}
\end{equation}
Furthermore, if we sum all Eqs. (\ref{EXTONE}) and also Eq. (\ref{EXTTWO}) and
use the resolution of the identity (\ref{POVM}),
we obtain formula for $\lambda$,
\begin{equation}
\lambda= \sum_{j=1}^N p_j\rho_j \Pi_j +a \sigma\Pi_0.
\label{LAMBDA}
\end{equation}
If we combine Eqs. (\ref{A}) and (\ref{LAMBDA})
then we can express $a$ and $\lambda$
solely in terms of $p_j$, $\rho_j$, and $\Pi_j$. This may be important,
for example, if we guess the optimal POVM and want to determine the
corresponding  Lagrange multipliers. The extremal Eqs. (\ref{EXTONE}) and
(\ref{EXTTWO}) constitute a generalization of the extremal equations for
optimal POVM for ambiguous quantum state discrimination that were
derived by Holevo and Helstrom \cite{Holevo_1973,Helstrom_QDET}.

We now provide simple sufficient conditions on the optimality of the
POVM. If the POVM satisfies the extremal Eqs. (\ref{EXTONE}) and
(\ref{EXTTWO}) and if the following inequalities hold:
\begin{eqnarray}
\lambda - p_j \rho_j & \geq & 0 , \quad j=1,\ldots, N,
\label{EXTINEQONE} \\
\lambda - a \sigma & \geq & 0,
\label{EXTINEQTWO}
\end{eqnarray}
then the POVM is the optimal one that maximizes the success rate $P_{\rm S}$
for a given fixed probability of inconclusive results $P_{\rm I}$.

To prove this statement we show that the Lagrange multipliers provide an
upper bound on the success rate and that this bound is saturated by the
POVM that satisfies Eqs. (\ref{EXTONE}) and (\ref{EXTTWO}).
From the definition of the success rate (\ref{SR}),
the inequalities (\ref{EXTINEQONE}) and  the normalization
(\ref{POVM})  we obtain
\begin{equation}
P_{\rm S} \leq \sum_{j=1}^N {\rm Tr}[\lambda \Pi_j] =
{\rm Tr}[\lambda(\openone-\Pi_0)] .
\label{RSRBOUNDONE}
\end{equation}
Now we use the inequality (\ref{EXTINEQTWO}) and finally we take
into account the constraint ${\rm Tr}[\sigma\Pi_0]=P_{\rm I}$.
We arrive at
\begin{equation}
P_{\rm S} \leq {\rm Tr}[\lambda] - a P_{\rm I}.
\label{RSRBOUNDTWO}
\end{equation}
This last inequality shows that $P_{\rm S}$ is limited from above by the
quantity that depends only on the Lagrange multipliers $\lambda$
and $a$ and also on the fixed $P_{\rm I}$. If the POVM $\Pi_j$
satisfies the extremal Eqs. (\ref{EXTONE}) and (\ref{EXTTWO}) then
this upper bound is reached, as can easily be checked.

We have thus established a simple criterion for checking of the
POVM optimality. Of course, we would like to derive the optimal POVM
$\Pi_j$ for given $p_j$, $\rho_j$ and $P_{\rm I}$. The analytical
solution to this problem seems to be extremely complicated.
Nevertheless, recently it was pointed out that one can solve this
kind of problems very efficiently numerically
\cite{Jezek_Rehacek_Fiurasek_2002}.
One possible simple and fruitful approach is to solve the extremal
equations by means of repeated iterations
\cite{Fiurasek_2001_POVM,Rehacek_Hradil_Fiurasek_Brukner_2001,%
Fiurasek_2001,Jezek_Rehacek_Fiurasek_2002,Jezek_2002,Rehacek_2002}.
In principle, one could
iterate directly Eqs.~(\ref{EXTONE}) and (\ref{EXTTWO}).
However, the POVM elements $\Pi_j$  should be positive
semidefinite Hermitian operators. All constraints
can be exactly satisfied at each  iteration step if the extremal
equations are symmetrized. First we express
$\Pi_j=p_j\lambda^{-1}\rho_j\Pi_j$  and combine it with its Hermitian
conjugate. We proceed similarly also for $\Pi_0$ and we get
\begin{eqnarray}
&&\Pi_j= p_j^2 \lambda^{-1} \rho_j \Pi_j \rho_j \lambda^{-1} , \quad
j=1,\ldots,N,
\label{EXTSYMONE}\\
&&\Pi_0=a^2 \lambda^{-1} \sigma \Pi_0 \sigma \lambda^{-1}.
\label{EXTSYMTWO}
\end{eqnarray}
The Lagrange multipliers $\lambda$ and $a$ must be determined
self-consistently so that all the constraints will hold.
If we sum Eqs. (\ref{EXTSYMONE}) and (\ref{EXTSYMTWO}) and take into
account that $\sum_{j=0}^N \Pi_j=\openone$, we obtain
\begin{equation}
\lambda=
\left[\sum_{j=1}^N p_j^2 \rho_j\Pi_j\rho_j+ a^2\sigma \Pi_0 \sigma
\right]^{1/2} .
\label{LAMBDASYM}
\end{equation}
The fraction of inconclusive results calculated for the POVM after the
iteration is given by
\begin{equation}
P_{\rm I}= a^2{\rm Tr}[\sigma \lambda^{-1}\sigma \Pi_{0}\sigma
\lambda^{-1}].
\label{PINCSYM}
\end{equation}
Since the Lagrange multiplier $\lambda$ is expressed in terms of $a$,
Eq. (\ref{PINCSYM}) forms a nonlinear equation for a single
real parameter $a$ (or, more precisely, $a^2$).
This nonlinear equation can be very efficiently solved by Newton's
method of halving the interval.
At each iteration step for the POVM elements, we thus solve the system of
coupled nonlinear equations (\ref{LAMBDASYM}) and (\ref{PINCSYM})
for the Lagrange multipliers. These self-consistent iterations work
very well and our extensive numerical calculations confirm that
they typically exhibit an exponentially fast convergence
\cite{Jezek_Rehacek_Fiurasek_2002}.

We note that the maximization of the success rate $P_{\rm S}$ for a fixed
fraction of inconclusive results $P_{\rm I}$ can also be formulated
as a semidefinite program. Powerful numerical methods developed for
solving this kind of problems may be applied. Here we will not
investigate this issue in detail and we refer the reader to the papers
\cite{Jezek_Rehacek_Fiurasek_2002,Eldar_Megretski_Verghese_2002,Eldar_2002}
where the formulation of optimal quantum-state discrimination
as a semidefinite program is described in detail.  Note also that
the semidefinite programming has recently found its applications in
several branches of quantum information theory such as
the optimization of completely positive maps
\cite{Audenaert_Moor_2002,Fiurasek_Iblisdir_Massar_Cerf_2002,
Verstraete_Verschelde_2002_A},
the analysis of the distillation protocols that preserve the positive
partial transposition \cite{Rains_2001},
or the tests of separability of quantum states
\cite{Doherty_Parrilo_Spedalieri_2002}.

\section{Maximal achievable relative success rate}

As the fraction of inconclusive results is increased the success rate
$P_{S}$ decreases. However, the relative
success rate ${P_{\rm RS}}$ grows until it achieves its maximum. If
$\{\rho_j\}_{j=1}^N$  are linearly independent pure states, then
$P_{\rm RS,max}=1$  because exact IDP scheme works and the unambiguous
discrimination is possible. Generally, however the maximum is lower than
unity. To find this maximum, we notice that in the limit
$P_{\rm I}\rightarrow 1$ the POVM element $\Pi_{0}$ must tend to the identity
operator. This means that at some point $\Pi_0$ becomes positive
definite operator and all its eigenvalues are strictly positive.
In that case, the extremal equation (\ref{EXTTWO}) can be satisfied if
and only if
\begin{equation}
\lambda= a \sigma.
\label{MAXIMUM}
\end{equation}

Since we are looking for some nontrivial solution to the extremal
equations with $P_{\rm I}<1$, at least one of the extremal Eqs.
(\ref{EXTONE})
must have a nontrivial solution $\Pi_{j}\neq 0$. This implies that
at least one of the operators $\lambda-p_j\rho_j$ must have one
eigenvalue $\mu$ equal to zero which implies that
\begin{equation}
{\rm det}[a\sigma-p_j\rho_j]=0
\label{DET}
\end{equation}
must hold at least for one of the states $\rho_j$. The optimal $\Pi_j$
is then proportional to the projector to the subspace spanned by
eigenvectors corresponding to the eigenvalue $\mu=0$.

The maximal attainable relative success rate is obtained if we
insert (\ref{MAXIMUM}) into (\ref{RSRBOUNDTWO}), take into account
that ${\rm Tr}[\sigma]=\sum_j p_j=1$ and re-normalize according
to Eq.~(\ref{RSR}),
\begin{equation}
{P_{\rm RS}}=a.
\label{RSRMAX}
\end{equation}
To determine the maximal ${P_{\rm RS}}$ we must find the maximal $a$  that
satisfies Eq. (\ref{DET}). Since $\sigma$ is positive definite it can be
inverted, and we can equivalently express the maximal $a$ as the maximal
eigenvalue of a Hermitian matrix,
\begin{equation}
a_j= p_j \max [{\rm eig}(\sigma^{-1/2}\rho_j \sigma^{-1/2})] .
\label{AEIG}
\end{equation}
The maximal ${P_{\rm RS}}$ is equal to the  largest $a_j$,
\begin{equation}
P_{\rm RS,max}= \max_j a_j.
\label{DOUBLEMAX}
\end{equation}
For qubits, Eq. (\ref{DET}) leads to quadratic equation for the
multiplier $a_j$ that can be solved analytically,
\begin{equation}
(a_j-p_j)^2=a_j^2 {\rm Tr}[\sigma^2]-2a_j p_j{\rm Tr}[\sigma \rho_j]
+p_j^2 {\rm Tr}[\rho_j^2].
\label{ajquadratic}
\end{equation}
It turns out that
the maximal ${P_{\rm RS}}$ depends only on the a-priori probabilities $p_j$,
the purities of the states ${\cal{P}}_j={\rm Tr}[\rho_j^2]$ and
the overlaps ${\cal{O}}_{jk}={\rm Tr}[\rho_j\rho_k]$. In this context
it is worth noting that it was shown recently that these parameters
of the quantum states can directly be measured without the necessity
to carry out a complete quantum state reconstruction
\cite{Filip_2002,Ekert_et_al_2002,Hendrych02}.

\section{Discrimination between two mixed qubit states}

We proceed to illustrate the methods developed in the present paper on
explicit example. We consider the simple yet nontrivial problem of
optimal discrimination between two mixed qubit states $\rho_1$  and
$\rho_2$. To simplify the discussion, we shall assume that the purities
of these states as well as the a-priori probabilities are equal,
${\cal{P}}_1={\cal{P}}_2=\cal{P}$, $p_1=p_2=1/2$.
The mixed states can be visualized as points
inside the Poincare sphere and the purity determines the distance of the
point from the center of that sphere. Without loss of generality, we can
assume that both states lie in the $xz$ plane and are symmetrically
located about the $z$ axis,
\begin{equation}
\rho_{1,2}=\eta \psi_{1,2}(\theta)+\frac{1-\eta}{2} \openone,
\label{RHO12}
\end{equation}
where the parameter $\eta$ determines the purity,
$\psi_j=|\psi_j\rangle\langle\psi_j|$ denotes a density matrix of
a pure state,
\begin{equation}
|\psi_{1,2}(\theta)\rangle= \cos\frac{\theta}{2}|0\rangle \pm
\sin\frac{\theta}{2}|1\rangle,
\label{PSI12}
\end{equation}
and $\theta\in(0,\pi/2)$.
From the symmetry it follows that the elements $\Pi_1$ and $\Pi_2$
of the optimal POVM must be proportional to the projectors
$\psi_1(\phi)$  and $\psi_2(\phi)$, where the angle $\phi\in(\pi/2,\pi)$
is related to the fraction of the inconclusive results. The third
component $\Pi_0$ is proportional to the projector onto state $|0\rangle$.
The normalization of the POVM elements can be determined from the
constraint (\ref{POVM}) and we find
\begin{equation}  \label{POVMEX}
  \begin{array}{rcl}
    {\displaystyle \Pi_{1,2}(\phi) } & {\displaystyle = } &
      {\displaystyle \frac{1}{2\sin^2(\phi/2)}\psi_{1,2}(\phi),} \\
    {\displaystyle \Pi_0(\phi) } & {\displaystyle = } &
      {\displaystyle \left(1-\frac{1}{\tan^2(\phi/2)}\right)
      |0\rangle\langle 0|.}
  \end{array}
\end{equation}
The relative success rate for this POVM reads
\begin{equation}
{P_{\rm RS}}=\frac{1+\eta\cos(\phi-\theta)}{2(1+\eta\cos\theta\cos\phi)}
\label{PRSEXAMPLE}
\end{equation}
and the fraction of inconclusive results is given by
\begin{equation}
P_{\rm I}= \frac{1}{2}\left(1+\eta\cos\theta\right)
\left(1-\frac{1}{\tan^2(\phi/2)}\right).
\label{PIEXAMPLE}
\end{equation}
The formulas (\ref{PRSEXAMPLE}) and (\ref{PIEXAMPLE}) describe
implicitly the dependence of the relative success rate $P_{\rm RS}$
on the fraction of the inconclusive results $P_{\rm I}$.
From Eqs. (\ref{A}) and (\ref{LAMBDA}) one can determine the Lagrange
multipliers $\lambda$ and $a$ for the POVM (\ref{POVMEX})  and check that
the extremal Eqs. (\ref{EXTONE}), (\ref{EXTTWO}), (\ref{EXTINEQONE})
and (\ref{EXTINEQTWO}) are satisfied
which proves that the POVM (\ref{POVMEX}) is indeed optimal.

The maximum ${P_{\rm RS,max}}$ (\ref{DOUBLEMAX}) is achieved if
the angle $\phi$ is chosen as follows,
\begin{equation}
\cos\phi_{\rm max}=-\eta\cos\theta.
\end{equation}
On inserting the optimal $\phi_{\rm max}$ back into
Eq. (\ref{PRSEXAMPLE}) we get
\begin{equation}
P_{\rm RS, max}=
\frac{1}{2}
\left[1+\frac{\eta\sin\theta}{\sqrt{1-\eta^2\cos^2\theta}}\right].
\label{RSREX}
\end{equation}
Making use of  Eqs.  (\ref{DOUBLEMAX}) and (\ref{ajquadratic}) we can express
the ${P_{\rm RS,max}}$ in terms of the overlap ${\cal{O}}_{12}$ and
the purity ${\cal{P}}$,
\begin{equation}
P_{\rm RS,max}=
\frac{1}{2}\left[1+\sqrt{\frac{{\cal{P}}-{\cal{O}}_{12}}%
{2-{\cal{P}}-{\cal{O}}_{12}}}\right].
\label{RSROP}
\end{equation}
If we calculate $\cal{O}$ and $\cal{P}$ for the density matrices (\ref{RHO12})
and insert them into (\ref{RSROP}) then we recover the formula (\ref{RSREX}).

\begin{figure}
\centerline{\psfig{figure=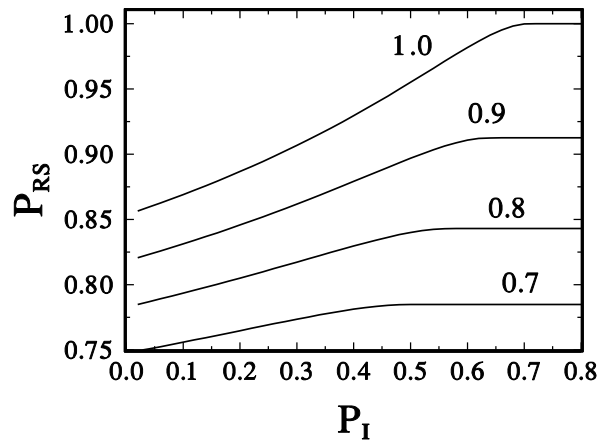,width=0.9\linewidth}}
\caption{Relative success rate ${P_{\rm RS}}$ versus the fraction of
inconclusive results $P_{\rm I}$ for the optimal discrimination
of two mixed states (\ref{RHO12}) with $\theta=\pi/4$ and four
different parameters $\eta=0.7$, $\eta=0.8$, $\eta=0.9$, and $\eta=1.0$.}
\label{TRADEOFF}
\end{figure}

The optimal POVM (\ref{POVMEX}) can be also obtained numerically.
We demonstrate feasibility of iterative solution of the symmetrized
extremal equations (\ref{EXTSYMONE}), (\ref{EXTSYMTWO}),
 (\ref{LAMBDASYM}), and (\ref{PINCSYM}) for mixed quantum states
(\ref{RHO12}) with the angle of separation $\theta=\pi/4$.
The trade-off of the relative success rate and the probability
of inconclusive results is shown in Fig.~\ref{TRADEOFF}
for various purities of the states being discriminated.
 For the given probability $P_{\rm I}$ of inconclusive results
and the given purity of the states the extremal equations are solved
self-consistently by means of repeated iterations.
The success rate ${P_{\rm RS}}$ is evaluated from
the obtained optimal POVM and re-normalized according to Eq.~(\ref{RSR}).
The numerically obtained dependence of $P_{\rm RS}$ on $P_{\rm I}$
is in excellent agreement with the analytical dependence following from
formulas (\ref{PRSEXAMPLE}) and (\ref{PIEXAMPLE}). Typically,
a sixteen digit precision is reached after several tens of iterations.
The trade-off curves  shown in Fig.~1 reveal the monotonous growth
of $P_{\rm RS}$ until the maximal plateau (\ref{RSREX}) is reached.

\section{Conclusions}
In conclusion, we have considered a generalized discrimination
scheme for mixed quantum states. The present scenario interpolates
between the Helstrom and IDP schemes. We allow for certain fixed fraction of
inconclusive results and maximize the success rate. We have derived the
extremal equations for the optimal POVM that describes the
discrimination device. The extremal equations can be very efficiently
solved numerically by means of the devised simple iterative algorithm or,
alternatively, by using the powerful techniques of semidefinite programming.

We have showed that the
relative success rate $P_{\rm RS}$ monotonically grows  as the fraction
of inconclusive results $P_{\rm I}$ is increased and at certain point it
reaches its upper bound $P_{\rm RS,max}$. For pure linearly independent
states this bound is $P_{\rm RS,max}=1$ which corresponds to the IDP
unambiguous discrimination scheme. For mixed states this bound is
in general lower than unity and we have derived a simple formula for it.

The present scheme may be important for quantum cryptographic schemes where
the receiver and/or eavesdropper want to discriminate nonorthogonal
states. Although these schemes are usually based on pure states,
in realistic cases the unavoidable noise and decoherence will reduce the
purity of these states and one would have to deal with mixed states.
Various modifications of our method can be suggested for such application.
For instance, if the involved states are in some sense asymmetric,
one may impose the condition that the probabilities
of inconclusive results or successful discrimination of the states $\rho_j$
should all be identical, and minimize the error rate with this additional
constraint. 

\begin{acknowledgments}

This work was supported by Grant LN00A015 and
Research Project CEZ: J14/98: 153100009 ``Wave and particle optics''
of the Czech Ministry of Education.
We would like to thank Miloslav Du\v{s}ek, Radim Filip, Zden{\v{e}}k Hradil,
and Jaroslav \v{R}eh\'{a}\v{c}ek for helpful discussions.
  
\end{acknowledgments}

\end{document}